\documentclass[doublecol]{epl2}
\usepackage{epsfig}
\usepackage{hyperref}
\usepackage{color}
\usepackage{amsmath}

\begin{document}
\title{Fermi arcs and pseudogap emerging from dimensional crossover at the
Fermi surface in La$_{2-x}$Sr$_x$CuO$_4$}
\shorttitle{Fermi arcs and pseudogap\ldots}

\author{P.~Lazi\'c\inst{1} \and D.~K.~Sunko\inst{2}}
\shortauthor{P.~Lazi\'{c}~and~D.~K.~Sunko}

\institute{                    
  \inst{1} R.~Bo\v skovi\'c Institute, - Bijeni\v cka cesta 54, HR-10000
Zagreb, Croatia.\\
  \inst{2} Department of Physics, Faculty of Science, University of
Zagreb, - Bijeni\v cka cesta 32, HR-10000 Zagreb, Croatia.
}

\abstract{
The doping mechanism and realistic Fermi surface (FS) evolution of
La$_{2-x}$Sr$_x$CuO$_4$ (LSCO) are modelled within an extensive ab-initio
framework including advanced band-unfolding techniques. We show that ordinary
Kohn-Sham DFT+U can reproduce the observed metal-insulator transition, when
not restricted to the paramagnetic solution space. Arcs are self-doped by
orbital charge transfer within the Cu--O planes, while the introduced Sr charge
is strongly localized. Arc protection and the inadequacy of the rigid-band
picture are consequences of a rapid change in orbital symmetry at the Fermi
energy: the material undergoes a dimensional crossover along the Fermi
surface, between the nodal (2D) and antinodal (3D) regions. In LSCO, this
crossover accounts for FS arcs, the antinodal pseudogap, and insulating
behavior in $c$-axis conductivity, all ubiquitous phenomena in high-T$_c$
cuprates. Ligand Coulomb integrals involving out-of-plane sites are
principally responsible for the most striking effects observed by ARPES in
LSCO.
}

\maketitle


Perovskite cuprate oxides have so far defied all efforts to understand their
superconductivity (SC). They are ionic crystals, typically antiferromagnetic
(AF) in the parent composition, which readily metallize upon doping. The
metallization is signalled in ARPES by the Fermi arcs~\cite{Yoshida12}, one of
the most scrutinized features of underdoped cuprates. On the other hand, the
question how the out-of-plane dopands introduce charge into the
planes~\cite{Mazumdar89} has been mostly sidelined, although the relationship
between charge and dopand concentration is not simple in
general~\cite{Obertelli92}. The critical issue is whether the arcs appear
because the doped charge is subject to 2D correlations, or the doping
mechanism imposes externally that the delocalized charge first appear on the
arcs.

The role of AF correlations in cuprate SC has similarly not been settled. In
the superexchange mechanism~\cite{Barriquand94}, AF is suppressed when the
bridging oxygen $2p^6$ orbital is opened with doping. In a striking series of
experiments in electron-doped cuprates~\cite{Tsukada05,Adachi13}, the $2p^6$
orbital was opened instead by direct lowering of the Cu--O crystal field
splitting $\Delta_{pd}$ through removal of the apical oxygen (T/T' effect).
The ensuing increase of the Cu--O covalency replaced the whole AF region of
the phase diagram by SC. At the same time, the LTT transition in La$_2$CuO$_4$
with Ba doping~\cite{Axe89} is evidence that SC is strongly suppressed by the
in-plane O$_x$--O$_y$ level splitting $\Delta_{pp}$~\cite{Barisic90}. Both
phenomena point to the interplay of covalent and ionic (crystal-field) effects
involving the in-plane oxygens in the emerging metallicity and SC in
cuprates.


Fermi surfaces in the cuprates cannot be modelled~\cite{Qimiao93} without
taking the in-plane oxygens into account either explicitly, in the Emery
three-band model~\cite{Emery87}, or implicitly~\cite{Pavarini01}, via a
next-nearest-neighbor ($t'$) extension to the $t$--$J$ model~\cite{Zhang88}.
The effective O$_x$--O$_y$ hopping $t_{pp}$ has to be rather large, even in
the electron-doped cuprates~\cite{Sunko07}, indicating significant
particle-hole (ph) symmetry breaking in the real materials, due to the
Cu--Cu-bridging O $2p_{x,y}$ orbital being much closer in energy to the Cu
$3d^9$ than to the $3d^8$ configuration, which is offset by the large energy
$U_d$ associated with triply-ionized Cu$^{3+}$. 

The observed importance of ph-symmetry breaking is significant,  because the O
$2p_{x,y}$ orbitals are not correlated to first order, pushing the material
away from the strong-coupling (ph-symmetric, ``one-band Hubbard'') limit.
However, a rigid-band tight-binding (TB) picture cannot fit FS evolution with
doping~\cite{Hashimoto08}, drawing attention to strong Coulomb interactions at
low energy scales.

This crossover between ionic and metallic limits is the main interest of the
present work. Focusing on LSCO, we give a unified description of the doping
mechanism, arc growth, and failure of rigid bands within a single DFT+U
framework. The 2D arcs appear as a 3D ionic background effect, without
invoking the many-body effects in the 2D metal, which may still be responsible
for the charge and spin order observed below the pseudogap temperature.

Calculations were done in Kohn-Sham DFT~\cite{Hohenberg64,Kohn65} with
projector-augmented wave (PAW) pseudopotentials~\cite{Blochl94,Kresse99},
using the VASP software package~\cite{Kresse93, Kresse96}. The functional used
was PBE~\cite{Perdew96} with +U
correction~\cite{Anisimov91,Anisimov93,Liechtenstein95,Dudarev98}. In all
calculations the cutoff of 520 eV for plane wave expansion was used together
with a well converged mesh of $8^3$ $k$-points~\cite{Monkhorst76} for the
smallest unit cell. The bandstructure was calculated along the high symmetry
directions for the 1x1x1 unit cell and for larger cells an advanced
zone-unfolding algorithm~\cite{Popescu12} was used as implemented in
Ref.~\cite{Medeiros14}. The size of the supercell limits the dopings we can
achieve. Here we focus on configurations $2\times 2\times 4$ and $2^3$, with
doping levels $6.25$\% and $12.5$\%, respectively.


\begin{figure*}
\begin{tabular}{cccc}
\includegraphics[width=4.3cm]{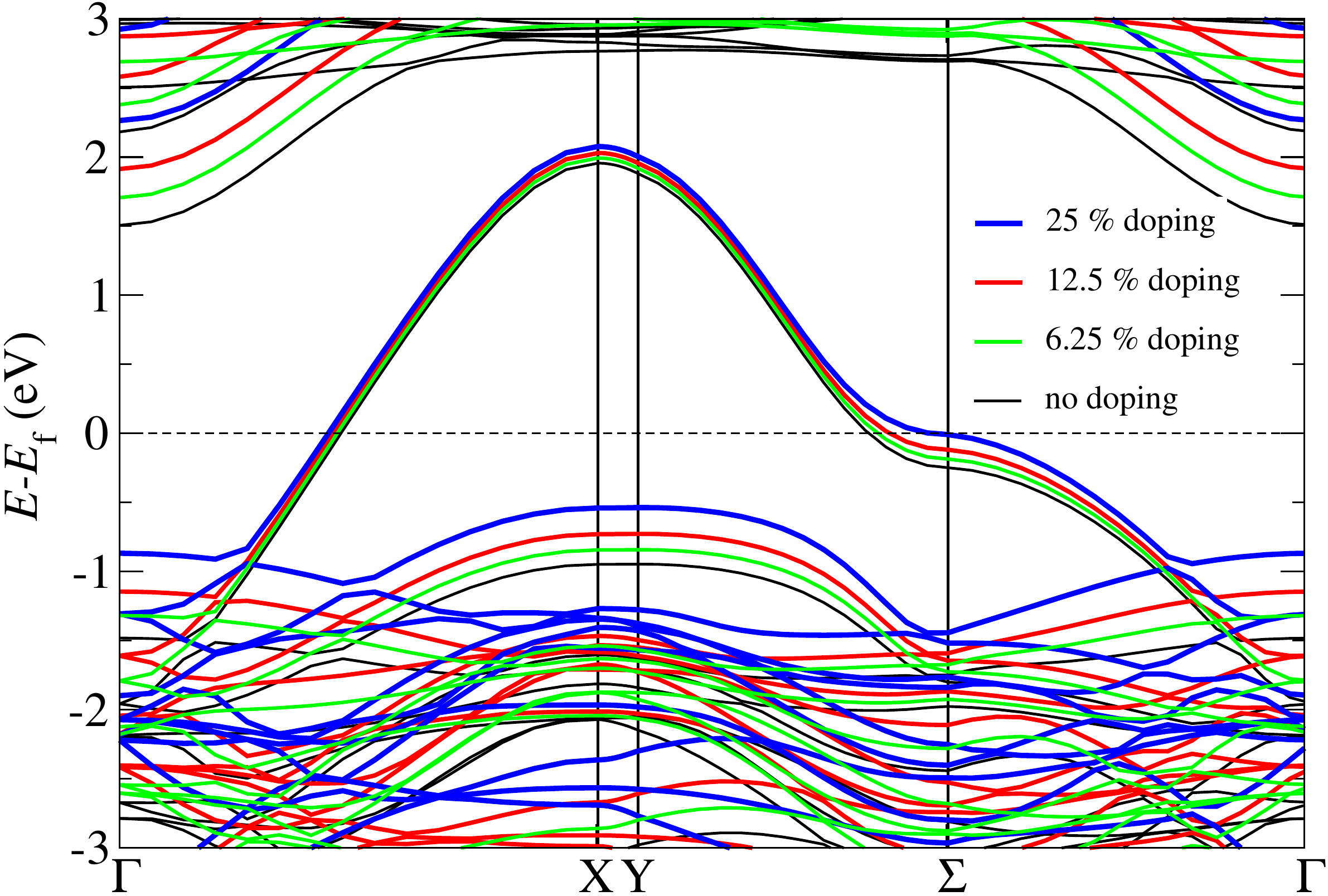}
&\includegraphics[width=4cm]{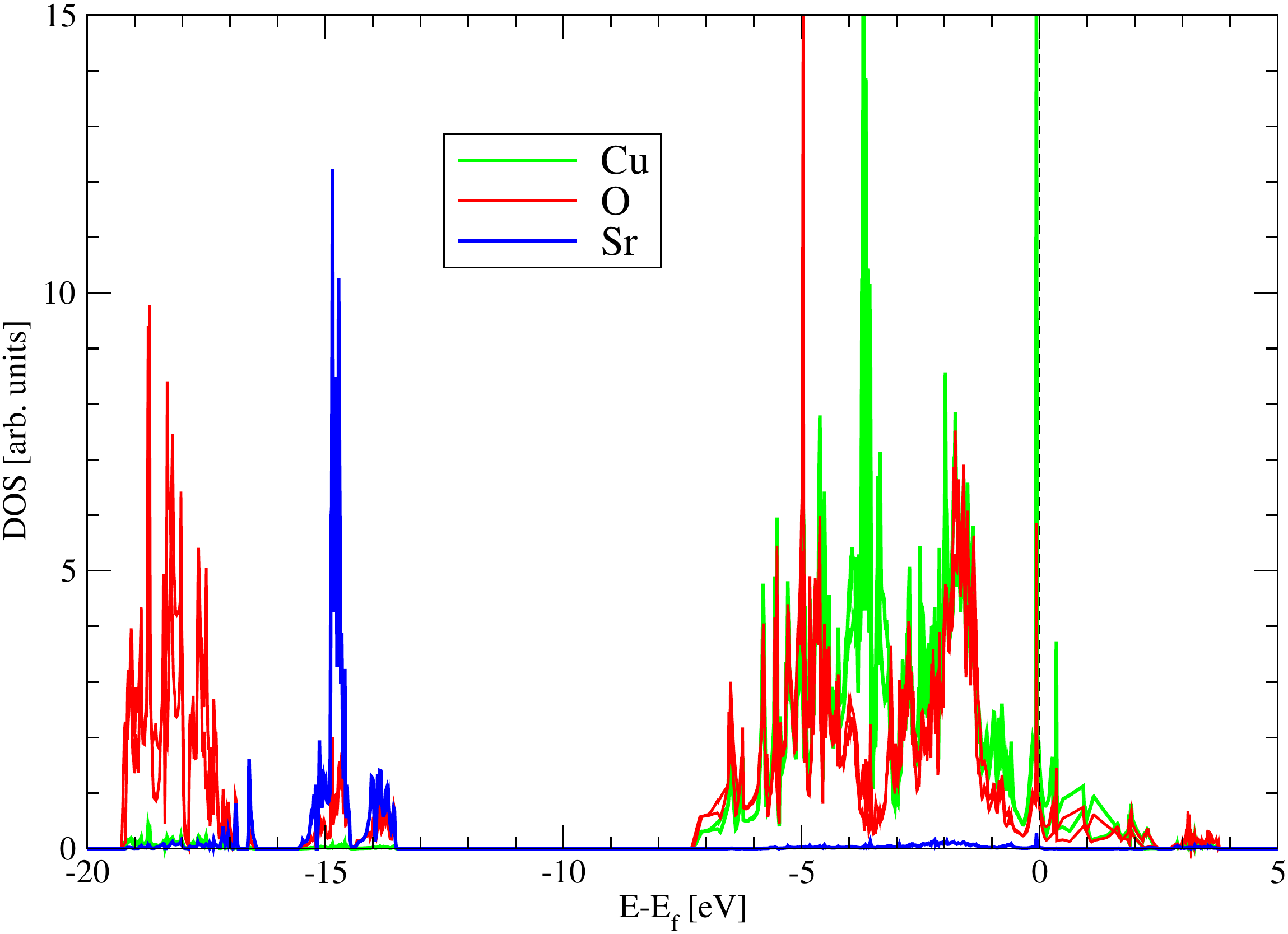}
&\includegraphics[width=4cm]{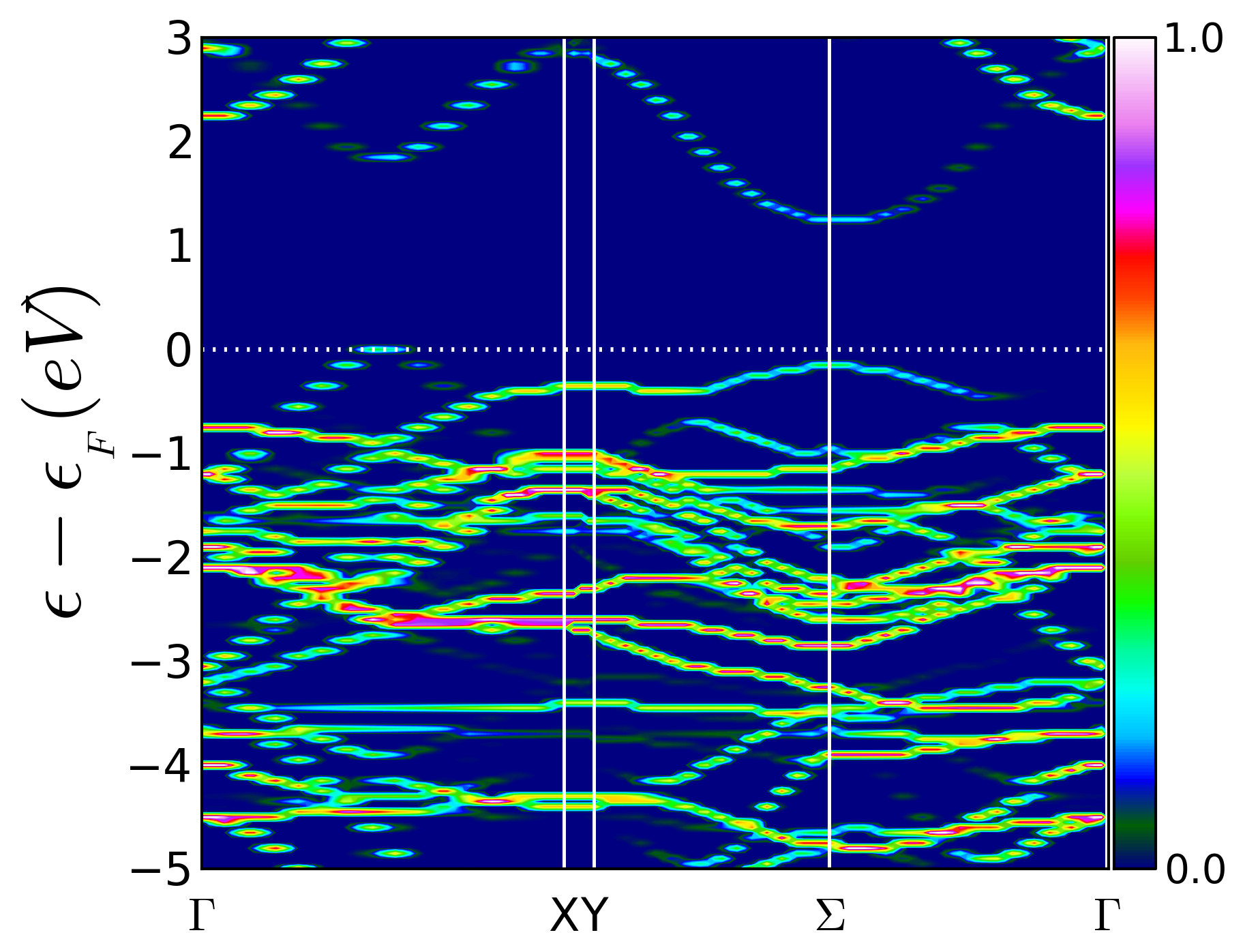}
&\includegraphics[width=4cm]{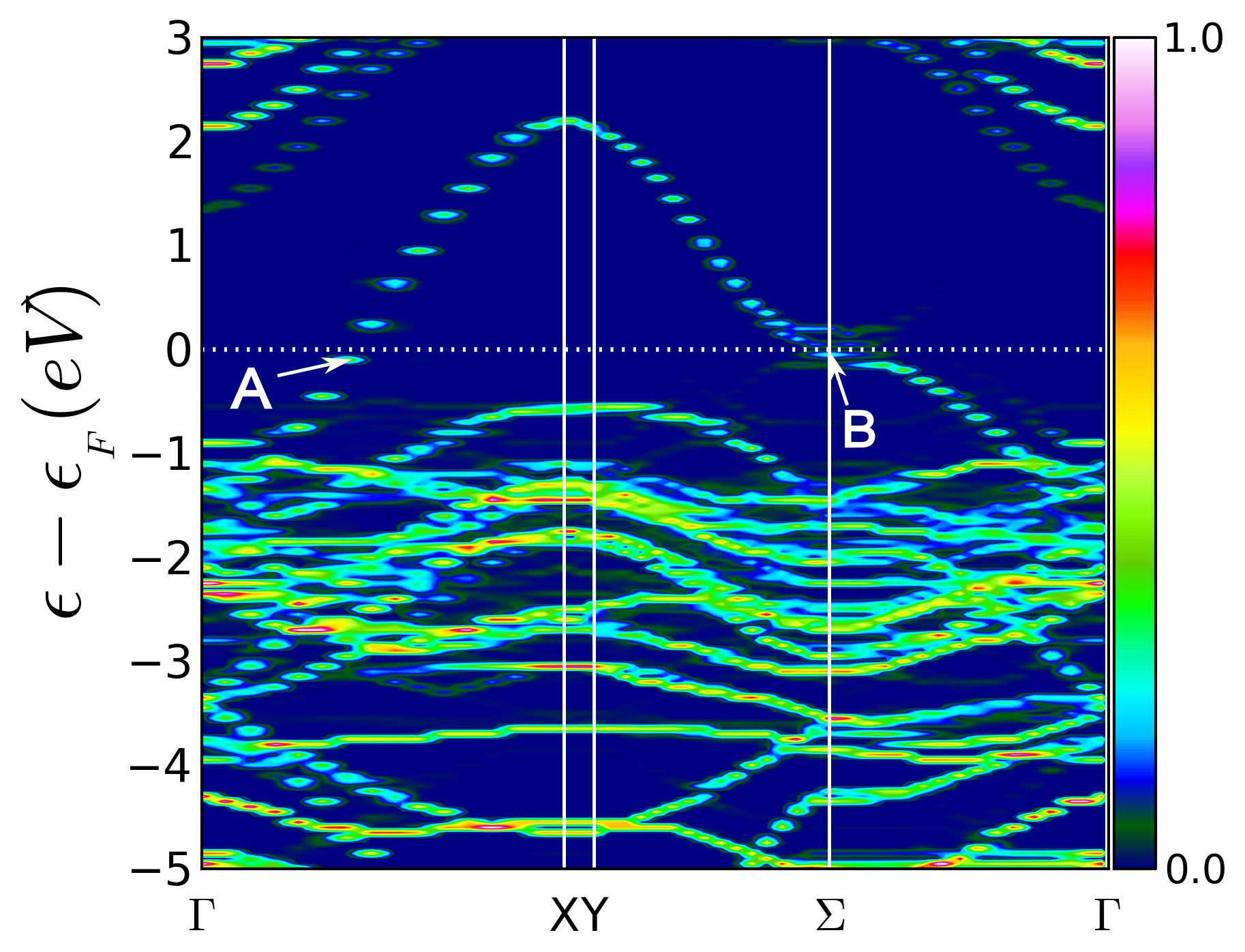}\\
(a) & (b) & (c) & (d)
\end{tabular}
\caption{(a) Gating simulation with paramagnetic starting point for four
characteristic fillings, thin to thick lines: $0$\%, $6.25$\%, $12.5$\% and
$25$\% doping, respectively. The shift in energy of the vH point ($\Sigma$) is
selectively enhanced. (b) Density of states (DOS) of the Sr dopand at 12.5\%
doping is concentrated around $E-E_f=-15$~eV. (c-d) Calculations for AF
starting point. Bands for $U_d=4$~eV at (c) zero doping and (d) 12.5\% doping.
The labels A and B in (d) are referred to in Fig.~\ref{fswf}.}
\label{pmevol}
\end{figure*}
Our first result appears already in a very simple calculation, in which doping
is simulated by the lack of an electron in a smeared positive background.  The
doping evolution is shown in Fig.~\ref{pmevol}a. The antinodal region clearly
shifts in energy more quickly than the nodal one, which means that the FS
evolution cannot be modeled by a rigid-band fit. However, the calculation
gives a paramagnetic metal (PM) at half-filling, contrary to experiment.

It has long been understood that the LDA approach cannot give an AF insulating
ground state at half-filling in the cuprates~\cite{Anisimov91}. An insulating
state in LSCO was obtained in the LDA+U approach~\cite{Anisimov92} without
gradient terms, but not a metal-insulator transition~\cite{Anisimov97}. Here
we revisit the issue within a generalized-gradient approximation (GGA+U) DFT
approach, separating the physical aspects from the technical. LDA+U gave a
band-gap which was too large, 2.3~eV~\cite{Anisimov92} compared to the
experimental 1.8~eV~\cite{Cooper90}. This fits with its general tendency to
overbind, due to neglect of gradient terms in the electron density, so that
sharply delimited charge distributions, good for binding, are unphysically
``cheap'' in kinetic energy. The GGA corrects this tendency by including the
gradient terms.

The paramagnetic metal at half-filling in GGA+U (Fig.~\ref{pmevol}a) is by
contrast a technical issue. Starting from scratch, a VASP calculation with
AF-initialized magnetic moments on copper atoms leads to a PM solution. We
avoid that solution by taking the converged PM solution as the initial one,
and restarting the calculation with initialized AF spin configuration on the
coppers. At half filling, our AF solution has a direct gap of 1.8~eV
(Fig.~\ref{pmevol}c), close to the experimental one, and significant energy
gain relative to the PM solution, which turns out to be less favorable
thermodynamically at all dopings. Even at $U_d=0$ a semimetallic solution with
zero density of states and a vanishing gap at $E_f$ is thermodynamically
preferred to the PM one. Our results do not depend qualitatively on the value
of $U$ at all, however when we choose the value $U_d=4$~eV, which fits the
enthalpy of formation~\cite{Hautier12}, the measured value of the band gap is
obtained as a prediction. An insulating AF ground state can be
modelled~\cite{Chen14} by DFT+U in charge-transfer insulators ($\Delta_{pd}\ll
U_d\to\infty$~\cite{Barriquand94}), even if Mott insulators
($\Delta_{pd}=U_d-\Delta_{pd}\to\infty$~\cite{Zhang88}) cannot be. Physically,
the bridging oxygens qualitatively prevent the correlation of neighboring
sites, which is unavoidable in one-band models of the Mott transition, as
manifested e.g. by the three-body hopping term in the $t$--$J$
model~\cite{Zhang88}. The oxygens open a significant alternative channel for
hopping even near half-filling~\cite{Ebrahimnejad14}.

Standard DFT+U is highly prone to gapping instabilities if allowed to break
local symmetries. When the relatively small supercell is allowed to relax
fully in both spins and atomic positions, translational repetition promotes
physical short-range order into unphysical long-range order, so a problem of
too little metallicity appears, rather than too much. Even slight technical
modifications of the DFT approach can promote gapping. Results closely
conforming to experiment are obtained in the present case if one breaks the
magnetic symmetry but does not allow atomic relaxation. (This means our
supercell is small, not that there is no atomic relaxation in the real
system.)

Introducing physical dopands now, we confirm strong
localization~\cite{Perry02} of the Sr charge. Although only 5--10\% of the
introduced hole charge reaches them, the planes screen the dopands
metallically: there is practically no induced charge beyond the two CuO$_2$
planes nearest to the Sr dopand. The local effect of the Sr atom is to
surround itself with negative charge, as if completing the \emph{orbital}
rather than the \emph{charge} configuration of La. It gets screened in turn by
a second layer of positive charge, so the material response to Sr is a
polarization (dielectric) effect carrying a net positive charge within $\sim
1.5$~\AA~of each dopand. We prove Sr charge localization in real space in
Fig.~\ref{pmevol}b, showing that the dopand DOS is narrowly concentrated in a
wide band gap far from $E_f$~\cite{Berlijn12}. Furthermore, we perfomed a
calculation at $1/8$ doping with twice-as-large unit cell and two physical Sr
dopands, which we placed in various positions. In all cases the distribution
of charge around each dopand was identical, indicating an extremely short
disorder correlation length.

There is no real-space Sr disorder in our supercell calculation. Disorder
effects on the metallic arc states are expected to be minor, because of the
short correlation length, Sr DOS isolation, and the orthogonality of the zone
diagonal to out-of-space orbitals (see below).

A $2,500^2$ unit-cell calculation in real space found a large amount of
disorder with percolating channels~\cite{Tahir-Kheli11}. Our physically
similar momentum-space calculation shows what ARPES can see of that real-space
picture. It seems to be a feature of out-of-plane Sr doping that metallicity
initially promotes disorder, or, conversely, that AF is more vulnerable to
disorder than metallization~\cite{Berlijn12}. Annealment by coherent hopping
similarly induces disorder in a solvable model of quantum percolation in the
plane~\cite{Sunko05}.

Two unfolded band structures are shown in Fig.~\ref{pmevol}c
and~\ref{pmevol}d, one at half-filling and the other at 1/8 doping. The former
has a large gap and an energy gain of $\sim 17$~meV per atom relative to the
PM one. The latter has a small (pseudo)gap only around the vH singularity, and
is only $\sim 1.5$~meV per atom better than the PM one. This partial gapping
is direct evidence of an arc-protection~\cite{Ferrero09} effect. In the
calculation the pseudogap is of mixed magnetic and orbital-disorder origin, as
discussed below.

\begin{figure}
\begin{center}
\begin{tabular}{cc}
\includegraphics[width=4cm]{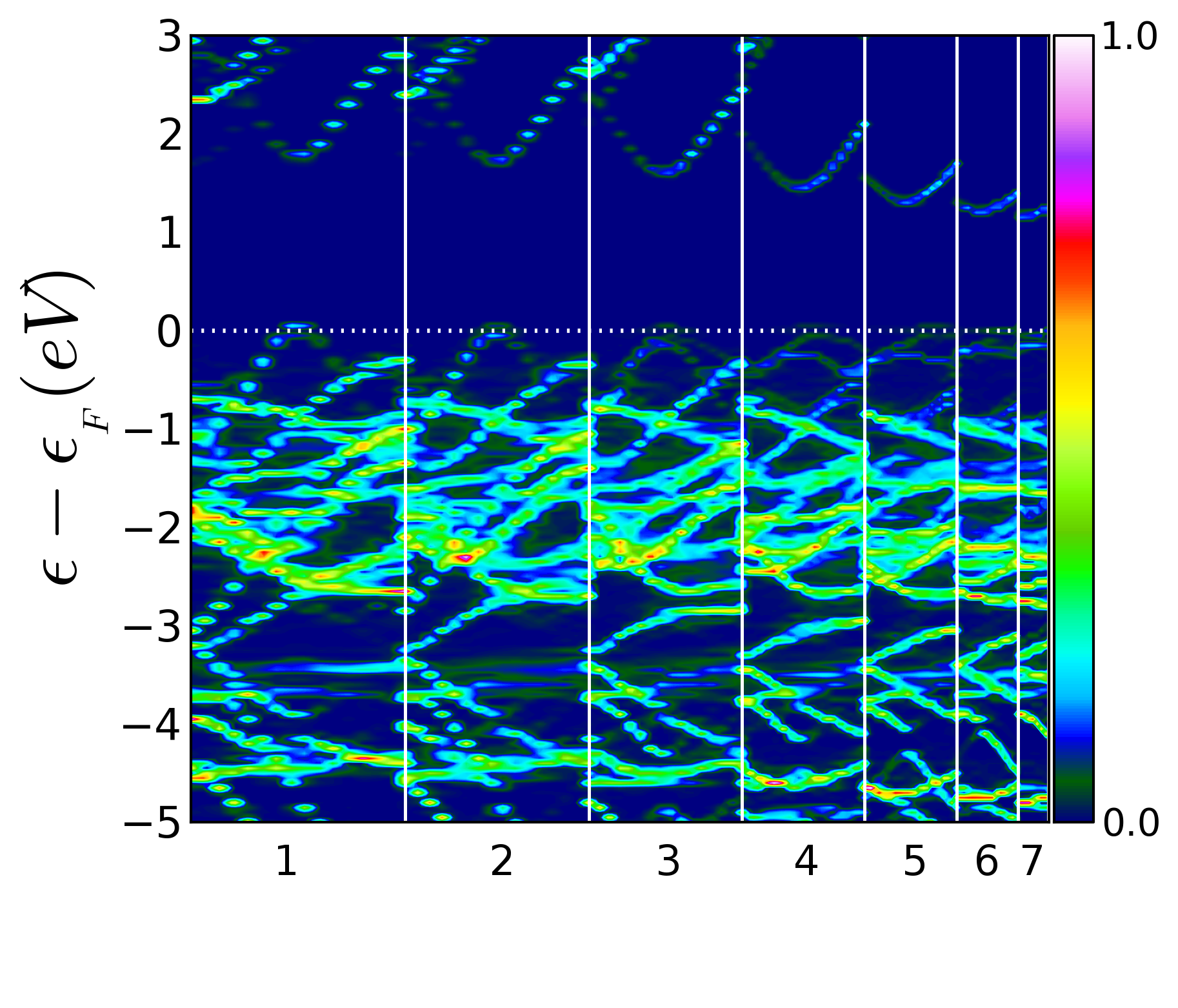}&
\includegraphics[width=4cm]{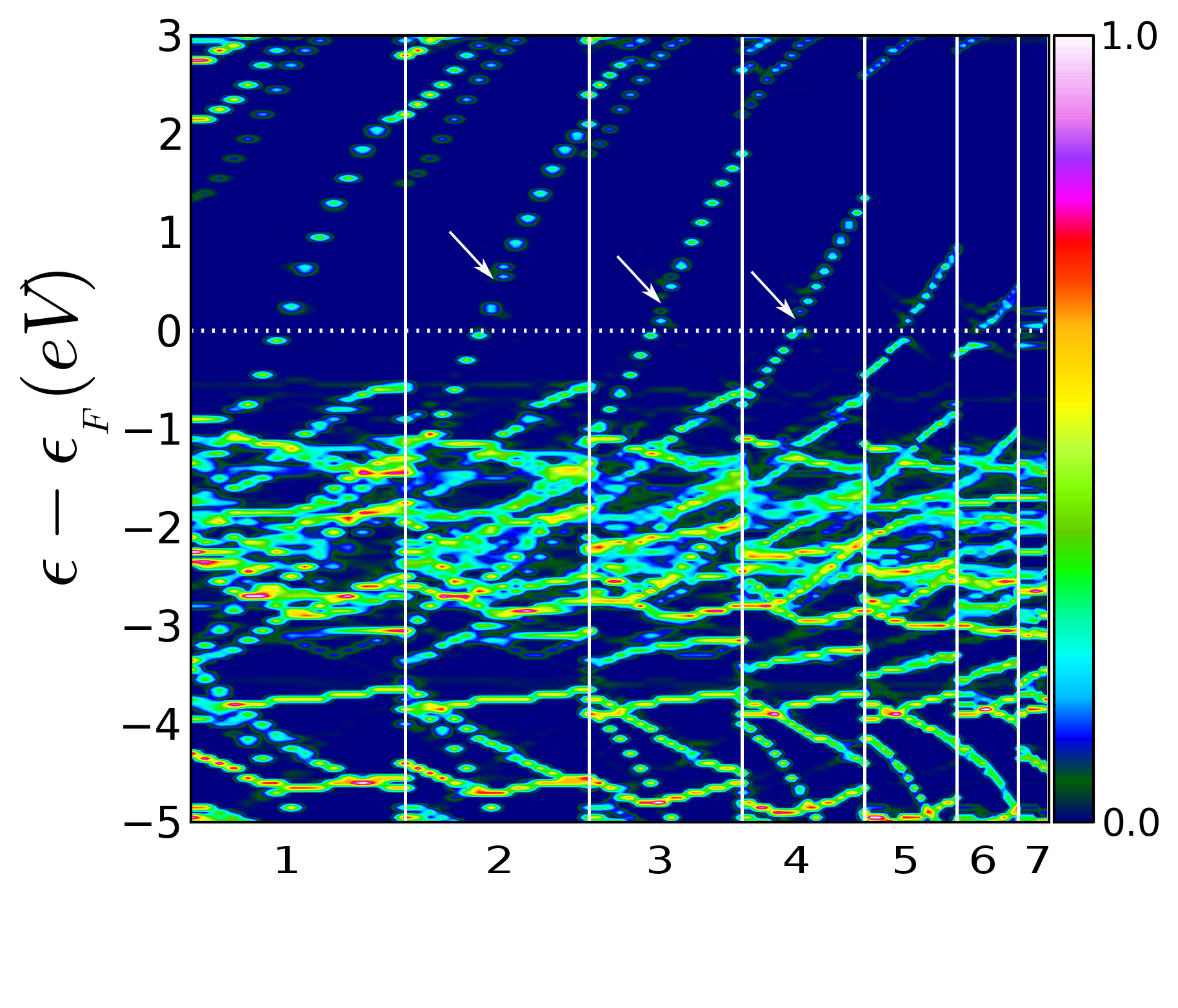}\\
(a) & (b)\\
\includegraphics[width=4cm]{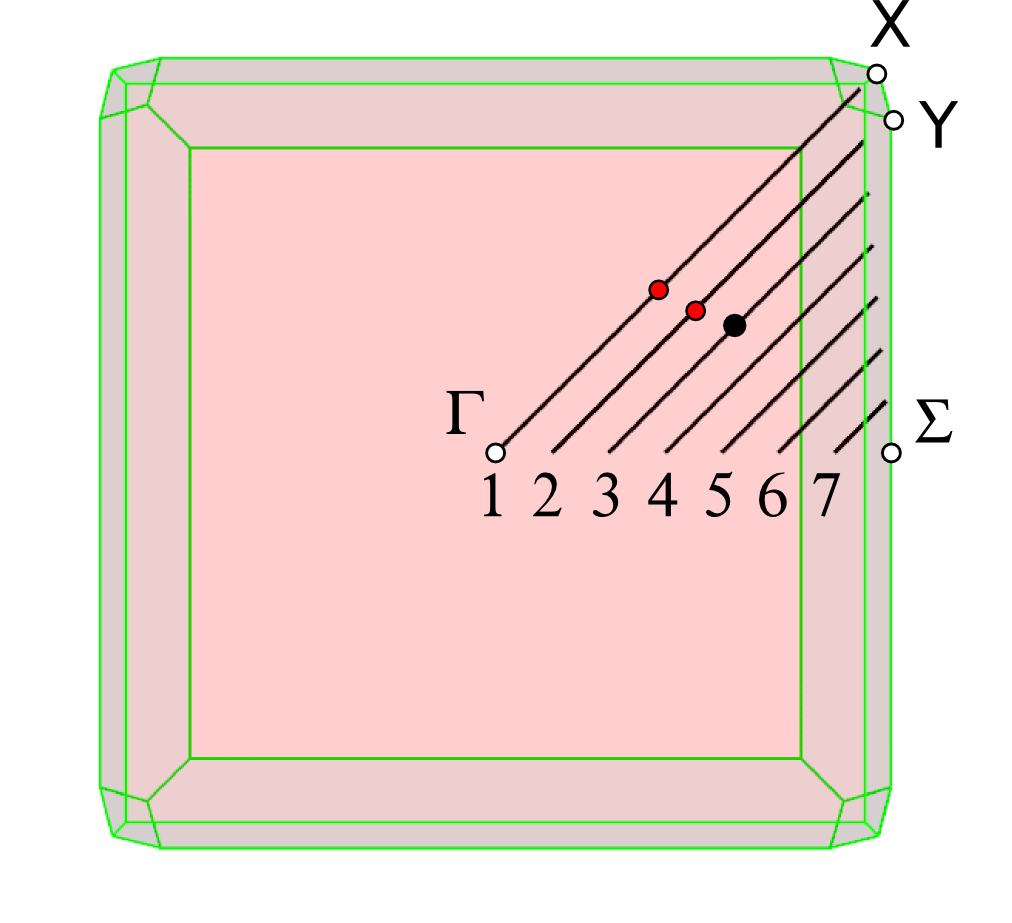} &
\includegraphics[width=4cm]{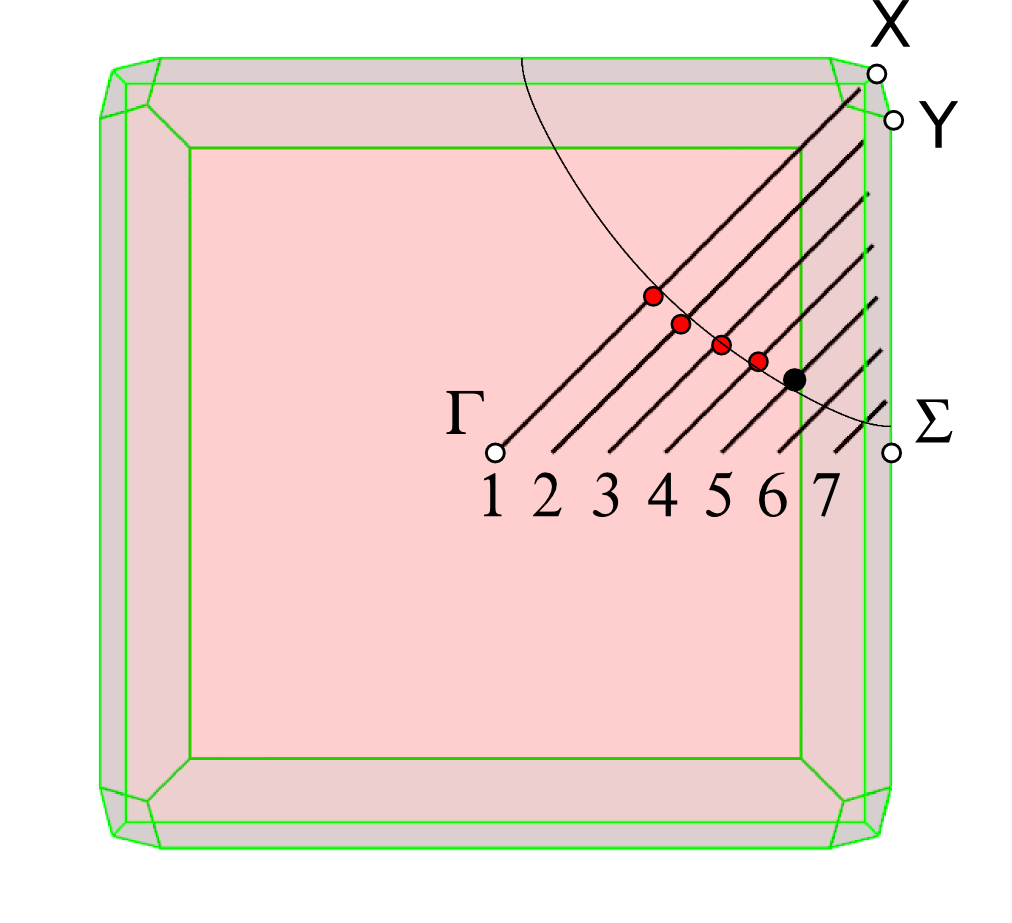}\\
(c) & (d)
\end{tabular}
\end{center}
\caption{Arc evolution. (a, b) Dispersions along BZ cuts from the nodal
towards the antinodal point (left to right), for (a) 6.25\% and (b) 12.5\%
doping. White arrows in (b) point to (pseudo)gap developing above the Fermi
energy. (c, d) Respective cuts in the BZ. FS crossing points are marked as red
dots, terminating with a black dot where the pseudogap is first observed. Thin
line in (d): best fit~\cite{Niksic14} to real data~\cite{Hashimoto08} at 15\%
doping.}
\label{arcevol}
\end{figure}

The arc evolution is further investigated in Fig.~\ref{arcevol}. We show cuts
in the BZ progressing from the nodal to the antinodal point, for 1/16 and 1/8
doping. The arc extends towards the end of the zone with increasing doping, as
expected from experiment~\cite{Yoshida12}. Significantly, it ends with a gap
opening rather abruptly: one can clearly see that the (pseudo)gap is
pre-formed above the Fermi energy, so that it is already open at the point
where it begins to straddle the FS. This is closely parallel to the real
situation in underdoped systems, where a step-like gap is found along the
FS~\cite{Lee07-1,He09}, as opposed to a ``d-wave'' gap near optimal doping. It
has also been observed in BSCCO that the step-like gap is an out-of-plane
effect~\cite{Okada08}. Thus the standard DFT+U method can cross the
metal-insulator (MI) transition in a manner closely resembling experiment,
including the formation and growth of metallic nodal arcs, when it is not
restricted to the PM solution space \emph{a priori}.

\begin{figure}
\begin{center}
\begin{tabular}{cc}
\includegraphics[width=3.5cm]{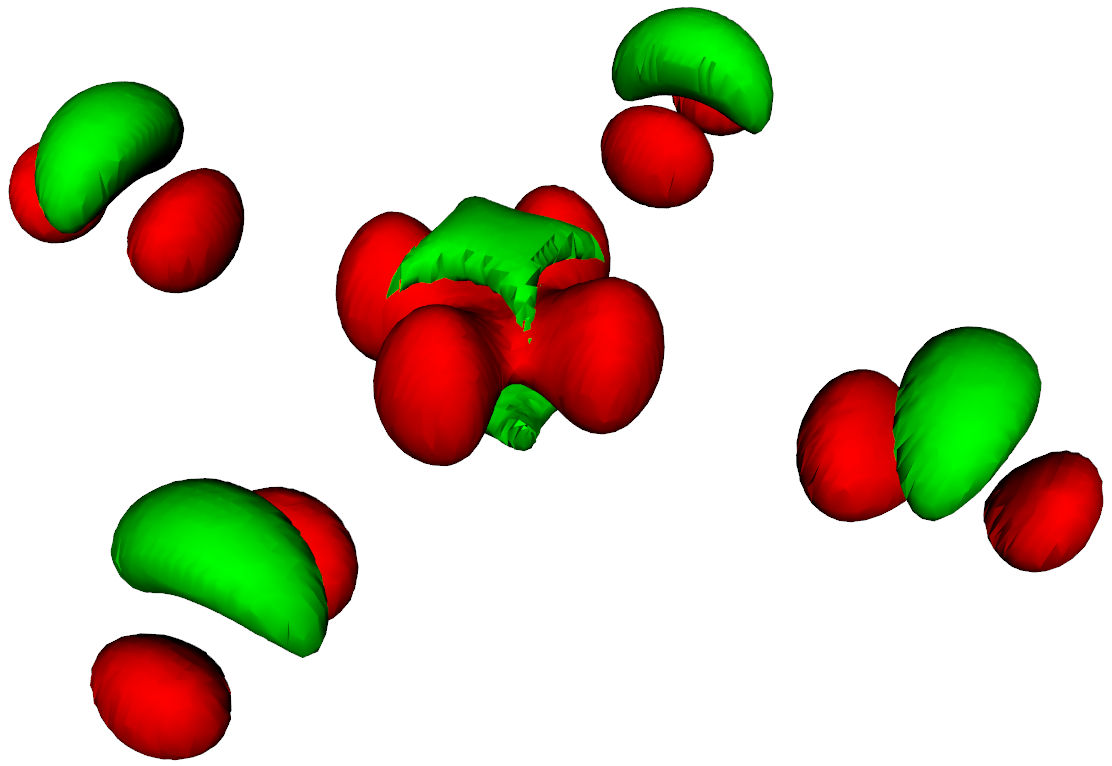}&
\includegraphics[width=3.5cm]{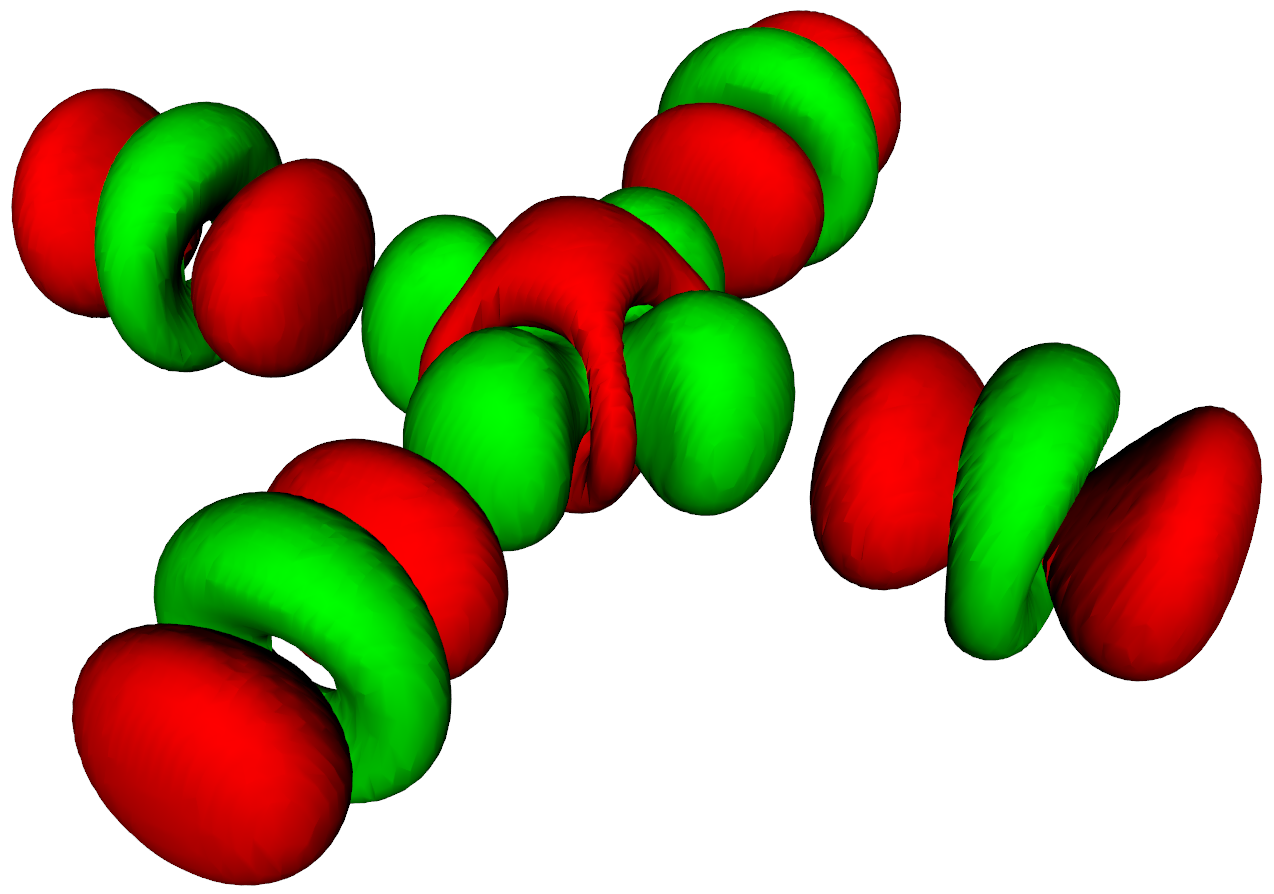}\\
(a) & (b) \\[1ex]
\multicolumn{2}{c}{\includegraphics[width=7cm]{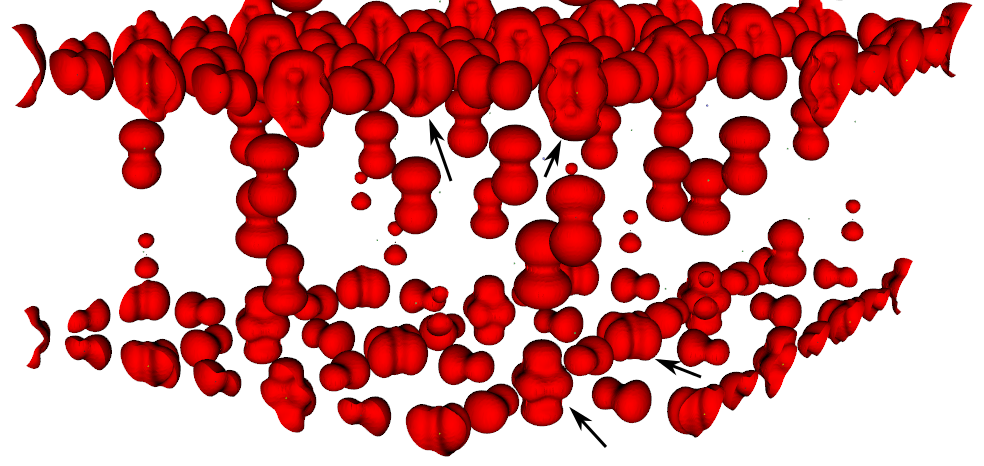}}\\
\multicolumn{2}{c}{(c)}
\end{tabular}
\end{center}
\caption{(a--b) Charge-difference isosurfaces relative to zero doping. Red
(green) surfaces are loci of constant $+1$($-1$)\%~$e$\AA$^{-3}$. (a) CuO$_4$
plaquette upon 1/8 Sr doping, calculated with PM starting point. It predicts
\emph{both} the Cu and O planar charge differences to be positive, relative to
the undoped case. (b) Charge-difference isosurfaces with AF starting point.
The planar charge difference upon 1/8 Sr doping is negative on Cu, positive on
O. (c) Isosurfaces of wave-functions at the vH point (labelled B in
Fig.~\ref{pmevol}d), at 1/8 doping, projected onto real space. The 3D
character is clearly visible. At least four different Cu orbital
configurations appear (black arrows), depending on the vicinity to the Sr
site. The corresponding wave-function at the nodal point (A in
Fig.~\ref{pmevol}d) is strictly planar (not shown).}
\label{fswf}
\end{figure}

Despite the Sr charge localization, the FS crossings in Fig.~\ref{arcevol}
conform closely but not perfectly with a Luttinger sum rule for a large
FS~\cite{Hashimoto08}, supporting the idea~\cite{Mazumdar89} that the majority
of the charge in the plane is delocalized \emph{in situ}, by the orbital
transition $\mbox{Cu$^{2+}+$O$^{2-}\rightleftharpoons\;$Cu$^{+}+$O$^{-}$}$.
The change in local Coulomb field when La$^{3+}$ is replaced by
Sr$^{2+}$ expels a Cu hole onto the oxygens, closing the Cu $3d^9$ orbital to
$3d^{10}$ and opening the O $2p^6$ orbital to $2p^5$~\cite{Mazumdar89}. Such
an \emph{ionic doping} mechanism has recently been directly
confirmed~\cite{Pelc15}, while the known alternatives have been disproved:
metallic alloying is excluded by the localization of the dopand
charge~\cite{Perry02}, also confirmed here, while semiconductor-like impurity
bands are excluded experimentally~\cite{Bozin05}.  We see its effect directly
by comparing charge transfers upon doping in the PM and AF solutions in
Fig.~\ref{fswf}a and \ref{fswf}b. In the former case, both the planar Cu and O
gain holes, like in the non-interacting TB models, while in the latter, Cu
\emph{loses} holes to the bridging oxygen, as expected in the ionic doping
mechanism.

The principal explanation for the above observations is given in
Fig.~\ref{fswf}c. It shows the wave-function content of the antinodal point in
the zone, marked B in Fig.~\ref{pmevol}d. One can observe at least four
different Cu orbital configurations (black arrows), while the wave function at
the nodal FS crossing (A in Fig.~\ref{pmevol}d) consists exclusively of the
planar O $2p_{x,y}$ and Cu $3d_{x^2-y^2}$ orbitals (not shown). The antinodal
pseudogap is characterized by strong orbital disorder, connecting to the
out-of-plane atoms via the Cu $4s$ and $3d_{x^2-y^2}$ orbitals. This result is
experimentally corroborated~\cite{Okada08} by manipulating the arc length with
out-of-plane disorder at \emph{fixed} level of in-plane doping. The 2D
metallic arcs are thus delimited by a 3D ionic background effect, obtained
here in a Fermi-liquid (FL) setting by construction. The coupling to the 3D
background explains why antinodal points move more quickly in energy with
doping, accounting for the failure of the rigid-band approach.

It was noted previously in stoichiometric LDA calculations that the PM
wave-function is 2D on the zone diagonal and 3D at the antinodal
point~\cite{Andersen95}, which was connected to the anisotropy in $c$-axis
transport~\cite{Ioffe98}. Here we confirm that result in a more realistic
pseudogapped setting with FS arcs at finite doping. This robustness of the
dimensional crossover protecting the arc is due to an  orbital symmetry
effect, which can be understood analytically. Out-of-plane orbitals can be
subsumed in an effective $4s$ Cu orbital, which with the usual three planar
orbitals~\cite{Emery87} makes the ensuing four-band model the minimal one with
chemically realistic values of TB parameters~\cite{Pavarini01}. The secular
polynomial of this model factorizes along the zone diagonal ($k_x=k_y$) in
such a way that the dispersion along the conducting band diagonal is uncoupled
from the Cu $4s$ orbital which is the physical conduit of out-of-plane effects
into the plane. This symmetry decoupling of the diagonal is exact, unlike the
\emph{parametric} decoupling~\cite{Pavarini01} of the planar dispersion for
small $t_{ps}/(\epsilon_d-\epsilon_s)$. It is the physical mechanism of arc
protection.

Interestingly, the O$_x$--O$_y$ splitting $\Delta_{pp}$ appearing in the LTT
tilt~\cite{Barisic90} breaks this symmetry. It couples the states on the
diagonal to out-of-plane orbitals via the Cu $4s$ state by a mixed
covalent-ionic term
$\left(\Delta_{pp}/2\right)^2(\epsilon_d-\epsilon_k)(\epsilon_k-\epsilon_s)$
in the secular polynomial of the four-band model. The Fermi-energy
($\epsilon_k=\epsilon_F$) distance from the Cu $3d_{x^2-y^2}$ orbital
$\epsilon_d$ measures the Cu--O covalency in the conduction band, while
relative to the Cu $4s$ orbital $\epsilon_s$ it measures the crystal-field
splitting between the (hole) valence and conduction bands. The observed strong
suppression of SC~\cite{Axe89} is thus related to the removal of arc
protection, lending support to the idea that SC in the underdoped cuprates
first develops on the arcs~\cite{Tanaka06}. One can interpret the T/T'
effect~\cite{Tsukada05,Adachi13} similarly: physical removal of the apical
oxygens is analogous to their removal by symmetry from the arc, cf.\
Fig.~\ref{fswf}, pushing the particular material in both cases towards the 2D
covalent limit~\cite{OSBarisic12}. Why this limit is so important for
high-T$_c$ SC remains an open question.

The observation of arc protection for interstitial-oxygen
doping~\cite{Lee07-1,He09,Vishik14} invites the conjecture that interstitial
sites are also hidden from the zone diagonal, which is plausible if the
(effective) Cu $4s$ orbital is again the principal connection of the plane
with the third dimension. We propose that a dimensional crossover along the FS
is a universal property of all materials in which a FS arc is observed.
Currently it seems to be a unique feature of the underdoped high-T$_c$
cuprates. (A recent report of arcs in an iridate~\cite{Kim14} may have an
alternative interpretation~\cite{Torre14}.)

Now we turn to the destruction of AF by doping. At every $k$-point in an
AF-gapped band, there is a superposition of two states of the ungapped band,
$\phi(k)$ and $\phi(k+Q_{AF})$. This superposition is manifested in the
real-space basis as strict staggering of magnetic moments for all the wave
functions in the zone, in particular for those on the zone-diagonal Fermi
surface (pocket) away from half filling. By contrast, we find real-space
staggering only in the antinodal wave-functions. At $1/8$ doping, the Cu
$3d_{x^2-y^2}$ antinodal staggered moment drops, from $0.55$ magnetons at
half-filling, to $0.07$ in the plane nearer the Sr dopand, and to $0.17$ in
the farther one, with some disorder in the values. (Similar reduced values of
spins were obtained in an earlier DFT calculation~\cite{Barbiellini08} with a
different approach to spurious metallicity.) For the arc states, staggering
relaxes into a longer wave, which is our model's manifestation of the
nearly-AF metal observed e.g.\ in $^{67}$Cu NMR~\cite{Moriya90}. Thus the arc
in the present work is not AF. As the small supercell favors AF, we
cannot trust the results quantitatively, however a more realistic calculation
including Sr disorder should further decrease the AF response.

The out-of-plane (3D) Coulomb effects which we study in LSCO have been
experimentally~\cite{Okada08} directly related to the arc length in BSCCO.
They appear at the antinodal region, concomitantly with remanent real-space AF
staggering in the small supercell. True Sr disorder, which we cannot treat due
to numerical limitations, should reduce the AF correlation length and increase
the orbital disorder at the antinodal point. The connection between AF and
dopand disorder is generally material-dependent~\cite{Andersen07}, with the
present model at one end of a continuum. The robust feature of the pseudogap
here is orbital disorder, whose large (ionic) energy scales make the step-like
pseudogap survive experimentally in the \emph{high-temperature}
limit~\cite{Razzoli13}, when smaller scales, like SC, have disappeared. At
intermediate temperatures still above T$_c$ we expect 2D metallic AF
correlations to influence the pseudogap~\cite{Niksic14}.

Importantly, Fermi arcs appear naturally in our calculation as an effect of
lattice-symmetry breaking. The periodically repeated large supercell maps onto
a small Brilloin zone, with a correspondingly complex Fermi surface due to
parent-compound bands folded over many times. In this small zone all Fermi
surfaces reach the zone edge by construction, because DFT+U is a one-body
model. Symmetry breaking occurs by band-unfolding, which projects the small
zone onto the reference parent-compound large zone, which cannot account for
the dopand positions. The net effect is that the states at the Fermi energy
in the small zone are ``left hanging'' in the large zone, i.e.\ one observes
them as a Fermi arc. The robustness of this symmetry-breaking mechanism is
attested by the fact that it already appears in our calculation with periodic
dopands. Further disorder in the dopand positions can only enhance the
effect. It explains the observation~\cite{Okada08} that different arc lengths
can correspond to the same concentration of mobile charge, because the length
of the arc in the large zone is not simply related to the Luttinger volume
enclosed by the multiple Fermi surfaces in the small zone. It also explains
why Fermi arcs are observed in the highly ordered mercury cuprates, where
local lattice effects of the interstitially doped oxygens in the mercury plane
on the far-away copper-oxide plane are expected to be slight. The
symmetry-breaking effect we describe could create an arc even if the
interstitial oxygens formed a perfect superlattice, as the Sr impurities do in
our calculation.

In a nutshell, arcs signify that the Coulomb effects of the dopands are not
averaged out at the level of a single unit cell in the copper-oxide plane. As
long as this is the case, dopand disorder is not expected to undo the effect
of lattice-symmetry breaking. This effect presumably decreases as the dopand
concentration increases toward  optimal, signalled by the arcs reaching the
edge of the large (CuO$_2$) planar Brillouin zone. The physical basis of this
mechanism is that the material response to doping remains dielectric between
copper-oxide planes in the underdoped region. The metallicity of the arcs is
due to their protection from the 3D Coulomb effects by orbital symmetry, which
appears to be a highly specific feature of the cuprates.

To conclude, ligand Coulomb integrals involving out-of-plane sites are
principally responsible for the most striking effects observed by ARPES in
LSCO. They gap the antinodal region extrinsically, without requiring strong 2D
electron-electron correlations, usually invoked in Fermi-arc descriptions
limited to the CuO$_2$ plane. Standard Kohn-Sham DFT+U is a FL theory by
construction, hence its ability to reproduce all the main features of the
nodal arc, as observed in ARPES, is evidence that the 2D metallic arc states
are a FL, protected by symmetry from the 3D ionic background. This result
agrees with recent observations of FL $T^2$ and $\omega^2$ laws in transport
and spectroscopic measurements~\cite{NBarisic12,Mirzaei12}. It does not
preclude further strong-coupling effects, either from copper on-site repulsion
or such as the symmetry breaking by $\Delta_{pp}$, which may still hold the
key to the SC mechanism. However, the observation of Fermi arcs does not by
itself imply that the nodal metal is not a Fermi liquid.

\acknowledgements Conversations with S.~Bari\v si\'c, A.~Fujimori,
S.~Mazumdar, G.~Nik\v si\'c, D.~Pavuna and E.~Tuti\v s are gratefully
acknowledged. We thank I. \v{Z}uti\'c and the Center for Computational
Research, University at Buffalo. P. L. thanks Paulo V. C. Medeiros, Sven
Stafstr\"om and Jonas Bj\"ork for help with their BandUP code for unfolding
the bandstructure. D.K.S. thanks the organizers for an invitation and a
stimulating time at the ECRYS-2011 Workshop on Electronic Crystals.

This work was supported by the Croatian Government under project
No.~119-1191458-0512 and by the University of Zagreb grant No.~202301-202353.





\end{document}